\documentclass{aa}
\usepackage{graphics,epsfig}

\def\simless{\mathbin{\lower 3pt\hbox
     {$\rlap{\raise 5pt\hbox{$\char'074$}}\mathchar"7218$}}}   %< or of order
\def\simmore{\mathbin{\lower 3pt\hbox
     {$\rlap{\raise 5pt\hbox{$\char'076$}}\mathchar"7218$}}}   %> or of order

\def\msun{{\rm M}_\odot}

\begin{document}

\title{Energy and time-lag spectra of galactic black-hole X-ray sources
in the low/hard state}

\subtitle{}

\author
{Pablo Reig\inst{1,3}, Nikolaos D. Kylafis\inst{2,3},
Dimitrios Giannios\inst{2}
}

\institute{
G.A.C.E, ICMUV, Universitat
de Valencia, E - 46071 Paterna-Valencia, Spain
\and University of Crete, Physics Department, P.O. Box 2208, 710 03,
Heraklion, Crete, Greece
\and Foundation for Research and Technology-Hellas, 711 10, Heraklion,
Crete, Greece
}

\authorrunning{Reig, Kylafis \& Giannios}
\titlerunning{Energy and time-lag spectra of black-hole systems
in the low state}

\offprints{pablo@physics.uoc.gr}

\date{Received: \\
Accepted: \\}

\abstract{
Most, probably all, accreting binaries that are believed to contain a
black-hole emit radio waves when they are in the low/hard state. Whenever
this radio emission has been resolved, a jet-like structure has become
apparent.  We propose that Compton upscattering of low-energy photons  in
the jet can explain both the energy spectra and the time lags versus Fourier
frequency observed in the low/hard state of black-hole systems.   The
soft photons originate in the inner part of the accretion disk. We have
performed Monte Carlo simulations of Compton upscattering
in a jet and have found that for a
rather wide range of values of the  parameters  we can obtain power-law
high-energy X-ray spectra with photon-number index in the range 1.5 - 2
and power-law time lags versus Fourier frequency with index $\sim 0.7$.
The black-hole source Cyg X-1 in the low/hard state is well described by
our model.
}

\maketitle

\section{Introduction} \label{intro}

The continuum X-ray spectra of black-hole candidates are generally
described by a soft component, normally modeled as a  multi-colour
black-body  component  (Mitsuda et al. 1984, but see Merloni et al. 2000)
and a power-law hard tail, which is thought to be the result of
Comptonization (see e.g. Sunyaev \& Titarchuk 1980; Hua \& Titarchuk 1995;
Titarchuk et al. 1997; Psaltis 2001). Based upon the presence or absence
of the soft component, the luminosity and spectral slope of the hard tail
and the different shapes of the noise components in the power-density
spectra, black-hole systems can be found in  at least two spectral
states:  The low/hard and the high/soft states. In the low/hard state the
soft component is weak or absent, whereas the hard tail extends to a few
hundred keV in the form of a power law with photon-number index in the
range 1.5 - 2. The power spectrum shows strong band-limited noise with a
typical strength of 20\% - 50\% rms and a break frequency below 1 Hz (van der
Klis 1995).

There is growing evidence that all X-ray binaries harboring a black hole
display radio emission when they are in the low/hard X-ray state (Fender
2001).  Inverse Compton scattering by relativistic electrons in a jet has
been proposed as a mechanism for the production of X-rays and
$\gamma$-rays in Active Galactic Nuclei (Begelman \& Sikora 1987;
Bednarek et al. 1996; Harris \& Krawczynski 2002) and X-ray binaries
(Band \& Grindlay 1986; Levinson \& Blandford 1996;  Georganopoulos et
al. 2002; Romero et al. 2002). The contribution of the synchrotron
emission from the jet to the hard X-rays might also be significant
(Markoff et al. 2001).

However, the Comptonization spectra cannot provide, by themselves, 
information about the geometry and the dynamics of the Comptonizing
electrons, i.e., one can obtain similar energy spectra from rather
different geometric configurations of the Comptonizing cloud (Titarchuk \&
Lyubarskij 1995). In addition to the energy spectra, time variability
information is needed (Hua et al. 1999).  Moreover, it has been observed
(Miyamoto et al. 1988) that the phase lags (not the time lags) between the
signals in two different X-ray energy bands are approximately constant
with Fourier frequency.  This cannot be explained if Comptonization takes
place in a hot, uniform, low-density plasma. A clever way to bring into
terms thermal Comptonization and timing behavior was to consider a
stratified spherical medium with density inversely proportional to radius
(Hua et~al. 1997; Kazanas et~al. 1997; Kazanas \& Hua 1999; Hua et~al. 1999).

Our proposal builds on the ideas presented in the above works.  The
relativistic electrons in the jet upscatter a fraction of the soft photons
emitted by the inner part of the accretion disk and this produces the
power-law energy spectrum. Power-law energy spectra with index in the
range 1.5 - 2 are produced with reasonable values of the parameters
describing the relativistic  electrons in the jet. In addition, we show for
the first time that, by considering a jet with density inversely
proportional to distance from the black hole, the nearly constant phase
lags with Fourier frequency can be reproduced. For an alternative
interpretation of time lags in black-hole candidates see Kotov et al.
(2001).

\section{The model}

\subsection{Characteristics of the jet}

We consider the simplest model that we can think, which reproduces the energy
spectrum and the time lags versus Fourier frequency that have been observed.
The characteristics of the assumed jet are the following:

i) We assume the speed of the relativistic electrons in the jet to be
constant.

ii) We assume that there exists a uniform magnetic field along the axis of
the jet (taken to be the $z$ axis) and that the velocity of the electrons
$\vec{v}$ consists of two components:  one parallel to the magnetic field
$\vec{v_{\parallel}}$ and one perpendicular to it $\vec{v_{\perp}}$. In
other words, the electrons in the jet are spiraling along the magnetic
field lines. This upscattering mechanism is reminiscent of thermal 
Comptonization  because the photons sample electrons with all possible
velocities (Titarchuk \& Lyubarskij 1995).

iii) We take the electron-density profile in the jet to be of the form
$$
n_{\rm e}(z)=n_0 (z_0/z)^p
\eqno(1)
$$
where $z$ is the vertical distance from the black hole and
$p$ is a parameter (which in this work is taken equal to 1) and
the parameters $n_0$ and $z_0$ are the electron
density and height at the base of the jet respectively.
If $H$ is the height of the jet, then the
Thomson optical depth along the axis of the jet is
$$
\tau_{\parallel} = n_0 \sigma_{\rm T} z_0 {\rm ln}(H/z_0),
\eqno(2)
$$
where $\sigma_{\rm T}$ is the Thomson cross section.

iv) Let $\pi r^2$ be the cross sectional area of the jet at height $z$.
Then, from the continuity equation  $\dot{M}~=~\pi r^2 m_{\rm p} n_{\rm e}(z)
v_{\parallel}$ and $p=1$, we obtain for the dependence of the radius $r$
on height $z$ that
$$r=(r_0^2 z/z_0)^{1/2},
\eqno(3)
$$
where  $r_0 = (\dot{M}/\pi m_{\rm p} n_0 v_{\parallel})^{1/2}$  is the cross
sectional radius of the jet at its base, $\dot{M}$ is the mass ejection
rate and $m_{\rm p}$ is the proton mass. Our jet is therefore focused, in
agreement with observations (Fender et al. 1999).  The half Thomson
optical depth of the jet at height $z$ is
$$
\tau_{\perp}(z) = n_0 \sigma_{\rm T} r_0 (z_0/z)^{1/2}.
\eqno(4)
$$

\subsection{The Monte Carlo code}

For our Monte Carlo code we follow Cashwell \& Everett (1959) and
Pozdnyakov et al. (1983).  A similar code was described in previous work
(Reig et al. 2001).  The procedure is as follows:

Photons from a blackbody distribution of characteristic temperature
$T_{\rm bb}$ are injected at the  center of the base of the jet with an upward
isotropic distribution. As the photons travel through the medium, they
experience Compton scatterings with the spiraling electrons.  If the
effective optical depth that they encounter is small ($\simless 1$), the
majority of the input photons  escape unscattered. If the effective
optical depth is moderate, then the photons random walk  through the
medium prior to escape and  on average gain energy from the circular
motion  (i.e., $\vec{v_{\perp}}$) of the electrons.  Such Comptonization
can occur everywhere in the jet.

If the defining  parameters of a photon (position,  direction, energy and
weight) at each stage of its flight are computed, then we can determine
the spectrum of the radiation  emerging from the  scattering medium and
the time delay of each escaping photon. The optical depth to  electron
scattering, the frequency shift and the new direction of the photons after
scattering are computed using the corresponding relativistic expressions.
The extra time of flight of each photon outside the jet is taken into
account in order to bring in step all the photons escaping in a given
direction.

\section{Parameter values}

The high-energy X-ray spectra of black-hole candidates in the low/hard
state are characterized by a power law with photon-number spectral index
in the range $1.5 \simless \Gamma \simless 2$ and a high-energy cutoff  in
the range $150 \simless E_{\rm cut} \simless 300$ keV (see e.g. Tanaka \&
Shibazaki 1996).   In addition, the soft thermal component, so prominent
in the high/soft state and interpreted as radiation from a cold accretion
disk, never totally disappears in the low/hard state. There is evidence
for a soft excess with blackbody temperatures $kT_{\rm bb}\sim 0.1-0.3$
keV (Baluci\'nska-Church et~al. 1995; Wilms et al. 1999). In particular,
for Cyg X--1 these parameters are $kT_{\rm bb} =0.13$ keV, $\Gamma =1.6$,
$E_{\rm cut}=160$ keV (Dove et al. 1998; di Salvo et al. 2001). Thus we
have considered a blackbody function with $kT_{\rm bb}=0.2$ keV as the
input source of soft photons in all the calculations presented in this
work.  Our conclusions are unchanged for $0.1 \simless kT_{\rm bb} \simless
0.3$ keV.

Observed time lags of the order of a fraction of a second, if interpreted
as light-travel times, require a height $H$ of the jet at least of the
order of $10^{10}$ cm.  In our calculations we have
taken $H = 2 \times 10^5 r_{\rm g}$,  where $r_{\rm g} = (GM/R)^{1/2}$ is the
gravitational radius of a black hole of mass $M$ and horizon radius $R$.
Thus, for a 10 $\msun$ black hole $H = 3 \times 10^{11}$ cm.

Both $E_{\rm cut}$ and $\Gamma$ depend on the velocity of the electrons in
the Comptonizing medium. In order to produce spectra similar to the ones
observed, we restricted the values of $v$ and $v_{\perp}$ to the ranges
$0.7c \simless v \simless 0.9c$ and  $0.3c \simless v_{\perp} \simless
0.5c$.  When another parameter in our  calculations is varied, we take $v
= 0.85 c$ and $v_{\perp} = 0.4 c$.

Another parameter of our model that affects the slope of the resulting
energy spectra is the width of the jet, since in a too narrow a jet too
few photons are upscattered.  The base of the jet was fixed at a
distance of $z_0=20 r_{\rm g}$ from the black hole, whereas $r_0$ was varied in the
range $75 r_{\rm g} \simless r_0 \simless 300 r_{\rm g}$.  When another
parameter in our calculations is varied, we take $r_0= 200 r_{\rm g}$.

In order to have enough scatterings to produce the desired power-law
energy spectra, we have taken $3 \simless \tau_{\parallel} \simless 15$,
with fixed value $\tau_{\parallel} = 10$ when another parameter is varied.
Note that the effective optical depth that the emitted photons see is
significantly less than this due to the high velocity of the electrons in
the jet.

Finally, given the high values of $v_{\parallel}$ invoked in our model, we
expect an angular dependence of the power-law X-ray spectra. From
observations of Galactic microquasars we know that the angle
between the observational direction
and the axis of the jet lies in the range
$70^{\circ}-85^{\circ}$ (Mirabel \& Rodr\'{\i}guez 1999). When a parameter
is varied in our model, we restrict ourselves to photons recorded in the
range $0.1 < \cos \theta < 0.3$ or $84^{\circ} > \theta > 72^{\circ}$,
where $\theta$ is the angle between the direction of motion of the jet and
the line of sight.

A thorough study of the dependence of the energy and time-lag spectra  on
the different parameters (including different density laws) will be
presented in a forthcoming paper (Giannios et al., in
preparation).

\section{Results and discussion}

In order to compute the energy spectra and the light curves we binned the
energy of the escaping photons, their travel time and their direction
cosine with respect to the axis of the jet.

\subsection{Energy spectra}

\begin{figure}
\mbox{}
\vspace{7.0cm}
\includegraphics{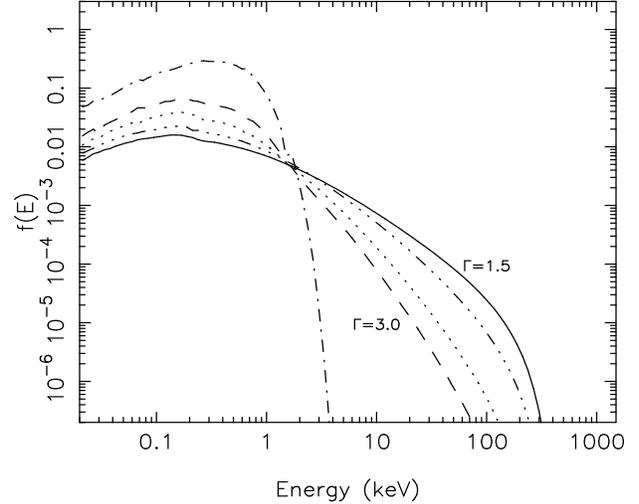}
\caption[]{Emergent photon-number spectra from our Monte Carlo simulations.
The plotted lines correspond to the values $\tau_{\parallel} = 0$
(dot-dashed),
3 (dashed), 5 (dotted), 10 (dot-dot-dot-dashed) and 15 (solid) of the vertical
optical depth.}
\label{energy}
\end{figure}

Figure \ref{energy} shows the emerging spectra from the jet for the fixed
values of the parameters discussed in section 3 and various optical depths
$\tau_{\parallel}=$ 0, 3,  5, 10 and 15.  All curves are normalized to
unity. The curve corresponding to $\tau_{\parallel} = 0$ gives the input
blackbody spectrum.  All spectra exhibit a soft, thermal-type component
and a hard, power-law type component with a high-energy cutoff.

\begin{figure}
\mbox{}
\vspace{8.0cm}
\includegraphics{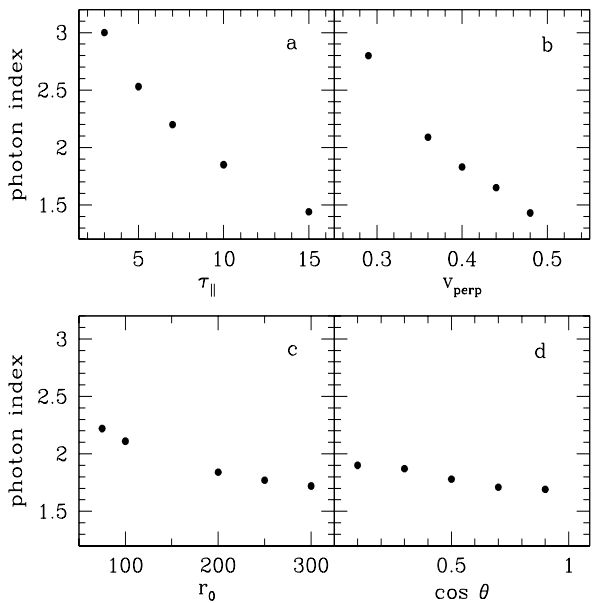}
\caption[]{Photon-number spectral index as a function of a) optical depth
$\tau_{\parallel}$, b) perpendicular component of the electron velocity
$v_{\perp}$, c) base radius of the jet $r_0$ and d) escaping angle $\theta$
with respect to the $z$ axis.}
\label{index}
\end{figure}

Figures \ref{index}$a-d$ show the photon-number spectral index $\Gamma$ of
the hard power-law tail in the energy range 10 - 100 keV as a function of
a) the optical depth $\tau_{\parallel}$, b) the perpendicular component of
the electron velocity $v_{\perp}$, c) the radius of the jet $r_0$ at its
base and d) the escaping angle $\theta$ with respect to the $z$ axis. For
intermediate optical depths, the photon-number index lies in the interval
$\sim 1.5 - 2$, which is similar to that observed in the low/hard state of
black-hole candidates (Tanaka \& Shibazaki 1996).
The thicker the medium the flatter the spectrum. The component $v_{\perp}$
must be a significant fraction of the total speed of the electrons in
order to have $1.5 \simless \Gamma \simless 2$. The larger $v_{\perp}$ the
flatter the spectrum. For the other two parameters, acceptable values of
the spectral index are obtained over a broad range of radii $r_0$ and
escaping angles~$\theta$.

%------------------------------------------------------------------------------
%------------------------------------------------------------------------------
%------------------------------------------------------------------------------
%------------------------------------------------------------------------------
%------------------------------------------------------------------------------
\begin{figure}
\mbox{}
\vspace{7.0cm}
\includegraphics{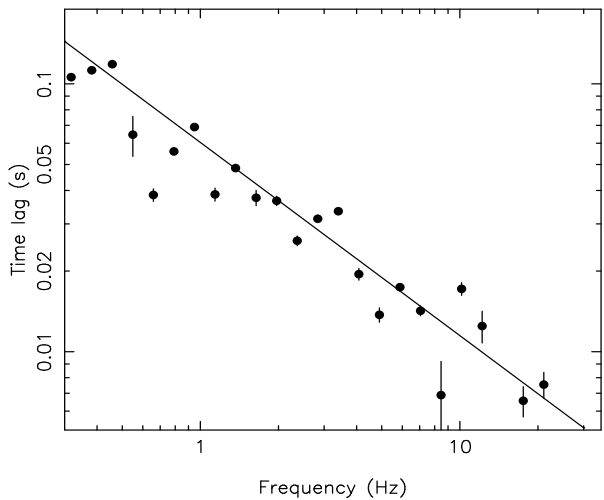}
\caption[]{Computed time lag as a function of Fourier frequency when the
fixed values of the parameters are used.  Positive lags
indicate that hard photons lag soft photons. The best-fit
power law is $\nu^{-0.7}$. }
\label{lagspec}
\end{figure}
%------------------------------------------------------------------------------

\subsection{Time lags}

The time of flight of all escaping photons was recorded in 8192 time bins
of duration 1/128 s each.  This time was computed by adding up the path
lengths traveled by each photon and dividing by the speed of light.  Then
we considered the light curves of two energy bands:  2 - 5 keV and
14 - 45 keV.  Following Vaughan \& Nowak (1997),
we computed the phase lag and through it the time lag between the two
energy bands as a function of Fourier frequency.

Figure~\ref{lagspec} shows the time lag as a function of Fourier
frequency for the fixed values of the parameters (section 3).
The data points have
been logarithmically rebinned for clarity.  The time-lag spectrum is
roughly represented by a power law $\nu^{-\beta}$
with index $\beta = 0.7$ in the frequency
range 0.3 - 30 Hz,  in excellent agreement with the results of Cyg
X-1 (Nowak et al. 1999).

\subsection{Energetics}

For the fixed values of the parameters
$\tau_{\parallel}=10$, $v=0.85c$, $v_{\perp}=0.4c$, $z_0 = 20 r_{\rm g}$,
$r_0=200 r_{\rm g}$ it is straightforward to show that the available kinetic
luminosity in the jet is much larger than the X-ray luminosity of
black-hole sources in the low/hard state.  Thus, the soft photons
steal only a small fraction of the energy in the jet with their upscattering.

\section{Conclusion}

We have shown that Compton upscattering of low-energy photons in a jet
can explain both the energy and time-lag spectra in black-hole X-ray
sources in the low/hard state. For reasonable values of the model
parameters we are able to reproduce the energy and time-lag spectra of
Cyg X-1 in the low/hard state.

\begin{acknowledgements}
PR is a researcher of the programme  {\em Ram\'on y Cajal} funded by the
University of Valencia and the Spanish Ministry of Science and
Technology. PR also acknowledges partial support from the Generalitat
Valenciana through the programme {\em Ajudes per a estades post-doctorals
en centres de fora de la Comunitat Valenciana}.
\end{acknowledgements}

\end{document}